\documentclass[conference]{IEEEtran} % Use the IEEE conference paper style

\usepackage[T1]{fontenc} % Proper font encoding
\usepackage{amsmath,amssymb,physics} % For advanced math symbols and equations
\usepackage{algorithm}
\usepackage{algpseudocode}
\usepackage{float}
\usepackage{graphicx} % For including graphics
\usepackage{cite} % For citation management
\usepackage{balance} % To balance the columns at the end of the document
\usepackage{qcircuit}
\usepackage{subcaption}
\usepackage{comment}
\usepackage{siunitx}
\usepackage{booktabs}
\usepackage{makecell}
\usepackage{multirow}
\usepackage{svg}
\usepackage{url}
\usepackage{xurl}
\usepackage[dvipsnames]{xcolor}
\usepackage{tabularx}
\usepackage{hyperref}
\usepackage[capitalize]{cleveref}
\definecolor{ionqorange}{HTML}{FF5000}
\hypersetup{
    colorlinks=true,
    linkcolor=ionqorange,    % Link color (internal links)
    citecolor=ionqorange,    % Citation color
    urlcolor=ionqorange      % URL color
}

\IEEEoverridecommandlockouts % For allowing the use of footnotes and more
\title{End-to-end Performance of Quantum-Accelerated Large-Scale Linear Algebra Workflows}
\author{
    \IEEEauthorblockN{
        Daiwei Zhu\IEEEauthorrefmark{1}, %\textsuperscript{\textsection},
        Miguel Angel Lopez-Ruiz\IEEEauthorrefmark{1},
        Fran\c cois-Henry Rouet\IEEEauthorrefmark{2},
        Claudio Girotto\IEEEauthorrefmark{1},
        Willie Aboumrad\IEEEauthorrefmark{1}, 
    }
    \IEEEauthorblockN{
        Robert Lucas\IEEEauthorrefmark{2},
        Ananth Kaushik\IEEEauthorrefmark{1},
        Martin Roetteler\IEEEauthorrefmark{1}
    }
    \medskip
    \IEEEauthorblockA{\IEEEauthorrefmark{1}IonQ Inc., 4505 Campus Dr, College Park, MD 20740, USA}
    \IEEEauthorblockA{\IEEEauthorrefmark{2}Synopsys, 675 Almanor Ave, Sunnyvale, CA 94085, USA}
}

\begin{document}

% Make title
\maketitle

\date{\today}
\begin{abstract}
Solving large-scale sparse linear systems is a challenging computational task due to the introduction of non-zero elements, or ``fill-in''. The Graph Partitioning Problem (GPP) arises naturally when minimizing fill-in and accelerating solvers. In this paper, we measure the end-to-end performance of a hybrid quantum-classical framework designed to accelerate Finite Element Analysis (FEA) by integrating a quantum solver for GPP into Synopsys/Ansys' LS-DYNA multiphysics simulation software. The quantum solver we use is based on Iterative-QAOA, a scalable, non-variational quantum approach for optimization. We focus on two specific classes of FEA problems, namely vibrational (eigenmode) analysis and transient simulation. 
We report numerical simulations on up to 150 qubits done on NVIDIA's CUDA-Q/cuTensorNet and implementation on IonQ's Forte quantum hardware. The potential impact on LS-DYNA workflows is quantified by measuring the wall-clock time-to-solution for complex problem instances, including vibrational analysis of large finite element models of a sedan car and a Rolls-Royce jet engine, as well as transient simulations of a drill and an impeller. We performed end-to-end performance measurements on meshes comprising up to 35 million elements. Measurements were conducted using LS-DYNA in distributed-memory mode via Message Passing Interface (MPI) on AWS and Synopsys compute clusters. 
Our findings indicate that with a quantum computer in the loop, amortized LS-DYNA wall-clock time can be improved by up to 14.6\% for specific cases and by at least 5.9\% for all models considered. These results highlight the significant potential of quantum computing to reduce time-to-solution for large-scale FEA simulations within the Noisy Intermediate-Scale Quantum (NISQ) era, offering an approach that is scalable and extendable into the fault-tolerant quantum computing regime.
\end{abstract}


\section{Introduction}

The efficient solution of large-scale, sparse linear systems remains a fundamental computational bottleneck in high-fidelity engineering simulations. In large-scale finite element analysis (FEA), matrices featuring tens to hundreds of millions of degrees of freedom must be factorized repeatedly. Although these matrices are inherently sparse, the factorization process introduces additional non-zero elements, a phenomenon known as fill-in, which significantly increases both memory requirements and computational overhead. Consequently, minimizing fill-in through matrix reordering via row and column permutations is critical for the performance of production multiphysics workflows. Modern solvers typically rely on Nested Dissection \cite{george1973nested} to manage this complexity which involves solving the Graph Partitioning Problem (GPP), a classic NP-hard optimization problem, repeatedly. This recursive strategy determines separators for sparse matrices represented as graphs, where the quality of each bipartition directly dictates the efficiency of the reordering, the extent of fill-in growth, and the total operation count required for factorization. Despite decades of development in multilevel heuristics such as METIS \cite{METIS}, graph partitioning remains an extremely challenging problem \cite{andreev2006gpp} and even marginal improvements in partition quality can yield substantial reductions in the runtime and memory footprint of industrial simulations.

Recent progress in quantum optimization has prompted investigations into whether quantum resources can enhance combinatorial subroutines within classical high-performance computing (HPC) environments. While hybrid quantum-classical approaches show promise, our prior work utilizing variational quantum imaginary time evolution (VarQITE) within the LS-DYNA pipeline revealed practical limitations regarding parameter optimization overhead and scaling difficulties with deeper circuits \cite{aboumrad2025accelerating}. To address these challenges, we introduce a strategy based on the Iterative Quantum Approximate Optimization Algorithm (Iterative-QAOA) \cite{lopez2025non} which is a warm-starting method for quantum optimization, see also \cite{Egger2021-kw, Willsch2022-cg, Sack2021-kf, Yu2022-rl,Yu2023-ix, Zhu2022-rl, Tate2023-rt, Cadavid2025-pm, Yu2025-bq} for similar approaches. 

Unlike standard QAOA, which requires tuning a vast set of variational parameters, Iterative-QAOA is non-variational and assigns circuit angles from a single scalar via a linear-ramp schedule, thereby eliminating the classical optimization loop. The algorithm's defining feature is an iterative warm-start mechanism: after circuit execution, measured bitstrings are scored by cost Hamiltonian energy, and a Boltzmann-weighted average of a low-energy subset is used to bias the initialization of the subsequent circuit. This progressively concentrates the quantum state on promising regions of the solution space without increasing parameter complexity, making the approach highly scalable and extensible to fault-tolerant hardware.

 Critically, we evaluate performance not through proxy graph-cut metrics, but by measuring end-to-end downstream simulation efficiency in terms of wall-clock time (WCT). By integrating quantum-generated partitions directly into LS-DYNA, we observe that Iterative-QAOA {\em consistently} identifies partitions that reduce sparse factorization fill-in, resulting in measurable runtime reductions. To ensure that quantum-enhanced partitioning provides a net gain in total WCT, the quantum execution time must be amortized over a sufficiently large classical FEA runtime. 
The gains reported in this work are therefore workflow- and
amortization-dependent: they presuppose a simulation regime in which a single reordering is reused across many subsequent solves.
Several simulation workflows involve performing matrix ordering once and reusing it over many subsequent steps. One is vibrational analysis, which involves computationally expensive eigenmode calculations to determine a structure's natural frequencies and mode shapes, preventing resonance, predicting dynamic responses, and enhancing structural durability. By simulating vibrations before physical testing, engineers identify weak points, reduce design failure, minimize prototyping costs, and prevent premature structural failures. Eigenmode calculations typically require solving a sequence of linear systems \(\left(K-\sigma_i M\right)u = f\), where \(K\) is the stiffness matrix, \(M\) is the mass matrix, and the scalars \(\sigma_i\)'s represent a sequence of so-called \emph{shifts}. These linear systems all have the same structure, so the ordering only needs to be computed once. Another candidate is transient analysis, which simulates time-dependent behavior under dynamic loading such as impacts or explosions or rapid temperature changes. Unlike steady-state analysis, it incorporates inertia, damping, and time-varying loads to predict how displacements, stresses, and strains change over time, crucial for safety and failure analysis. Assuming there are no severe deformations, or changes in contact, or element deletion (e.g., fracture), the same initial ordering can be used throughout the simulation. To further maximize gains, the quantum algorithm is applied only to the first level of Nested Dissection—the stage with the highest impact on reordering quality—while subsequent sub-problems are resolved using the classical LS-GPart heuristic. 

\begin{table}[!htb]
    \centering
    \caption{Summary of the problem instances analyzed in this paper, categorized by analysis type. Vertices and edges are expressed in millions (M). The D in the Mesh type denotes dimension. All models were used with permission.}
    \begin{tabular*}{\columnwidth}{@{\extracolsep{\fill}}lrrrr}
        \toprule
         & \multicolumn{2}{c}{Vibrational Analysis} & \multicolumn{2}{c}{Transient Analysis} \\ 
        \cmidrule(lr){2-3} \cmidrule(lr){4-5} \\[-1em]
                  &    SedanCar & JetEngine & Impeller &  Drill \\ 
        \midrule
        Vertices  &           5.9M &     34.9M &     7.0M &   3.6M \\
        Edges     &          55.3M &    870.6M &   182.5M & 101.2M \\
        Mesh type & Shells, solids &    Solids &   Solids & Solids \\
        Topology  &           2.5D &      2.5D &     2.5D &     3D \\ 
        Origin    &    ARUP, NHTSA & Rolls-Royce & Rolls-Royce & \makecell[r]{Predictive\\Engineering}\\
        \bottomrule
    \end{tabular*}
    \label{tab:problem_instances}
\end{table}

We apply this methodology to a diverse suite of industrial finite element models, including a car model, an impeller, a drill component, and a jet engine assembly. As shown in \cref{tab:problem_instances}, these models represent the distinct simulation regimes of transient dynamics and vibrational (modal) analysis. The SedanCar model was provided by ARUP and is based on an open source mesh of a Honda Accord~\cite{NHTSA}. JetEngine and Impeller are proprietary models provided by Rolls Royce, Inc. Drill is a proprietary model provided by Predictive Engineering, Inc.  Across these heterogeneous workloads, the quantum algorithm identifies partitions that systematically reduce factorization costs and accelerate end-to-end computation, yielding improvements  of up to about $14.6\%$ in total time-to-solution relative to the classical LS-GPart partitioner, and a consistent improvement of at least $5.9\%$ across all models considered in this study. To support this level of scaling, we utilize an in-house continuous problem-space reduction framework that adaptively maps large partitioning instances onto quantum hardware with a tunable number of qubits. This strategy preserves key structural features of the original graph while providing flexible control over quantum resources, enabling the application of Iterative-QAOA to graphs containing up to 10,000 nodes—a scale encountered in practical industrial configurations.

The remainder of this article is structured as follows. \cref{sec:background} establishes the theoretical framework by reviewing the formulation of graph partitioning as a Hamiltonian energy minimization problem. In \cref{sec:iterativeqaoa}, we describe the Iterative-QAOA algorithm. \cref{sec:methods} describes the integration pipeline of this quantum approach into LS-DYNA and defines the performance metrics used to benchmark our results against the classical LS-GPart heuristic solver. Experimental results, derived from both noiseless simulations and executions on IonQ hardware, are presented and analyzed in \cref{sec:results}. Finally, \cref{sec:future-outlook} discusses ongoing research trajectories and offers an outlook on the future development of hybrid quantum-classical methodologies.

\section{Graph Partitioning Problem Formulation}\label{sec:background}

We consider a vertex- and edge-weighted graph $G = (V,E)$. A bipartition of $V$ consists of two disjoint subsets whose union equals $V$. In the context of multilevel Nested Dissection as used in LS-DYNA, we seek balanced bipartitions that minimize edge cuts while satisfying a nodal weight constraint. Let $v_i$ denote vertex weights, $w_{ij}$ edge weights, and let $\Omega = \sum_i v_i$ be the total vertex weight. We require that the weight of each part does not exceed $(\tfrac{1}{2}+\nu)\Omega$, where $\nu$ is a prescribed imbalance tolerance (set to $\nu = 0.05$ in this work).

Introducing binary decision variables $x_i \in \{0,1\}$ to encode partition membership, the cut objective can be written as
\begin{equation}
\sum_{(i,j)\in E} w_{ij}(x_i + x_j - 2x_i x_j).
\end{equation}

To avoid explicit inequality constraints in the quantum formulation, we incorporate balance through a quadratic penalty term
\begin{equation}
P(x) = \left(\sum_i v_i x_i - \frac{\Omega}{2}\right)^2.
\end{equation}

This yields the unconstrained quadratic objective
\begin{equation}
C(x) = \sum_{(i,j)\in E} w_{ij}(x_i + x_j - 2x_i x_j)
+ \lambda P(x),
\end{equation}
where $\lambda > 0$ is a hyper-parameter controlling the trade-off between cut minimization and balance. The resulting problem is a Quadratic Unconstrained Binary Optimization (QUBO) instance.

Following standard constructions, we map the QUBO to a diagonal cost Hamiltonian $H_C$ acting on $n = |V|$ qubits by replacing each binary variable $x_j$ with the number operator
\begin{equation}
\hat{N}_j = \frac{1}{2}(I - Z_j),
\end{equation}
where $Z_j$ is the Pauli-$Z$ operator acting on qubit $j$. By construction,
\begin{equation}
H_C \ket{x} = C(x)\ket{x}.
\end{equation}

Thus, minimizing the graph partitioning objective is equivalent to preparing low-energy states of $H_C$. Additional derivations and implementation details can be found in our prior work~\cite{aboumrad2025accelerating}.

\section{The Iterative-QAOA algorithm}\label{sec:iterativeqaoa}

Solving the GPP reduces to sampling low-energy states of the cost Hamiltonian $H_C$. We do this with Iterative-QAOA~\cite{lopez2025non}, a non-variational QAOA variant that circumvents the costly classical parameter optimization limiting standard QAOA in large-scale settings.

At the core of Iterative-QAOA is a predetermined angle schedule that replaces per-instance variational fitting. Rather than optimizing $2p$ circuit parameters, where $p$ is the number of circuit layers, the algorithm uses a linear-ramp schedule (LR-QAOA)~\cite{Kremenetski2023-vy, Montanez-Barrera2025-im} to assign all angles from a single scalar $\Delta$, as follows:
\begin{align}
    \gamma_k = \frac{k}{p}\,\Delta, \qquad
    \beta_k  = \frac{p-k+1}{p}\,\Delta, \qquad k=1,\dots,p.
    \label{eq:lr_schedule}
\end{align}
With these angles fixed, the depth-$p$ QAOA circuit
\begin{align}
    \ket{\psi_p(\boldsymbol{\gamma},\boldsymbol{\beta})}
    = \prod_{k=1}^{p} e^{-i\beta_k H_M} e^{-i\gamma_k H_C}\ket{\psi_{\mathrm{init}}},
    \label{eq:qaoa_ansatz}
\end{align}
is executed from a product initial state $\ket{\psi_{\mathrm{init}}}$, where $H_M = \sum_j X_j$ is a transverse-field mixer. The circuit is measured to collect a set of bitstrings $\{\mathbf{x}_j\}$.

Measurement outcomes are scored by their cost Hamiltonian energy $E(\mathbf{x}_j) = \mel{\mathbf{x}_j}{H_C}{\mathbf{x}_j}$ and ranked through a Boltzmann distribution
\begin{align}
    P(\mathbf{x}_j) = \frac{e^{-\beta_T E(\mathbf{x}_j)}}{\sum_i e^{-\beta_T E(\mathbf{x}_i)}},
\end{align}
governed by an inverse temperature hyperparameter $\beta_T$. This energy-weighted ranking strongly amplifies the contribution of low-cost solutions. For each qubit $q$, a signed bias $m_q = \sum_j P(\mathbf{x}_j)(-1)^{x_{q,j}}$ condenses this collective signal into a directional preference. The initialization for the next circuit execution is then constructed as
\begin{align}
    \ket{\psi_{\mathrm{init}}^{(j+1)}} = \bigotimes_{q=1}^{n}\Bigl(\sqrt{1-\rho_q}\,\ket{0}+\sqrt{\rho_q}\,\ket{1}\Bigr),
    \label{eq:init_state}
\end{align}
where $\rho_q = \tfrac{1}{2}(1 - \eta m_q)$ and $\eta\in\{-1,+1\}$ selects the polarization direction. Following Ref.~\cite{Egger2021-kw}, the mixer is adapted so that $|\psi_{\mathrm{init}}^{(j+1)}\rangle$ remains an eigenstate of $H_M$. Repeating this execute-evaluate-reinitialize cycle progressively tilts the quantum state toward low-energy regions of the cost landscape, without any additional variational degrees of freedom beyond the single scalar $\Delta$.

\section{Methods}
\label{sec:methods}

\subsection{Overview of the end-to-end workflow}

\begin{figure}[t!]
    \centering
    \includegraphics[width=\columnwidth]{
    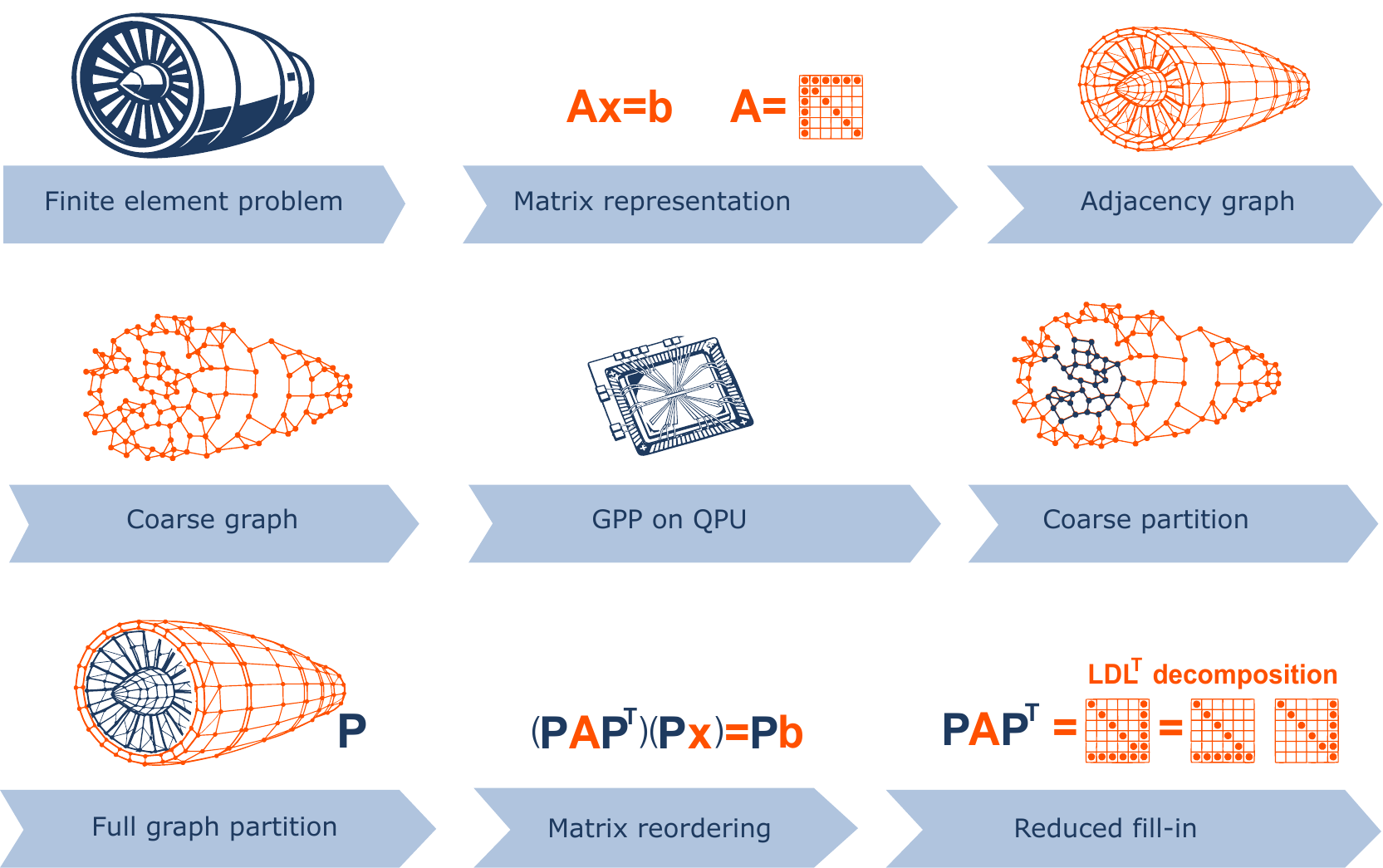}
    \caption{End-to-end quantum-accelerated linear algebra workflow. Starting from the mesh/adjacency graph of the sparse system, the original graph is coarsened to a hardware-matched size, the coarse graph partitioning problem (GPP) is solved on the QPU (Iterative-QAOA), and the resulting coarse graph bi-partition is lifted to a full-resolution graph partition. The lifted partition defines the separator/ordering used downstream for matrix reordering and sparse factorization.}
    \label{fig:cover_figure}
\end{figure}

In the context of finite element simulations, solving the sparse linear system $Ax = b$ is a central computational task, where the ordering of the matrix $A$ directly dictates the volume of fill-in and the overall cost of sparse factorization, such as $LDL^T$. To optimize this process, LS-DYNA constructs a permutation $P$ via Nested Dissection, a method in which vertex separators recursively bisect the adjacency graph of $A$ into smaller subgraphs. This generates a reordered matrix $PAP^T$ designed to reduce fill and accelerate both numerical factorization and back-substitution. In this study, we frame the selection of these separators as a balanced bi-partitioning objective on a weighted graph, minimizing the cut cost subject to a prescribed nodal weight balance tolerance. \cref{fig:cover_figure} illustrates the entire procedure and is similar to the procedure we have used in our prior publication \cite{aboumrad2025accelerating}. Our methodology utilizes a modified build of LS-DYNA that can export coarse subgraphs and ingest external partitions back into its internal Nested Dissection partitioner. The full adjacency graphs for our industrial benchmark instances (\cref{tab:problem_instances}) contain millions of edges and vertices, and they vastly exceed the capacity of current quantum hardware. To address this, LS-DYNA is first employed to generate and extract level-specific (level here refers to the level of Nested Dissection) coarse graphs restricted to 10,000 nodes. This 10,000-node baseline is consistent with production-level configurations of the LS-DYNA solver and provides a standardized platform for comparing our quantum GPP approach against the internal LS-GPart heuristic.

However, as 10,000 nodes still surpass the limits of near-term QPUs, we introduce an internal, hardware-matched coarsen–solve–lift loop. We utilize spectral coarsening (see \cref{SI}) to further reduce the graph to a flexible size compatible with the available QPU or simulator budget. This secondary coarsening stage aggregates vertex and edge weights so that the simplified graph structure accurately reflects the separator quality of the original 10,000-node instance. We then solve the coarse GPP using the Iterative-QAOA algorithm, which produces a distribution of candidate bi-partitions. The optimal candidate is selected and "lifted" back to the 10,000-node resolution, where it is returned to LS-DYNA to define the full-resolution separator. For multi-level Nested Dissection, this recursive coarsen–solve–lift procedure is applied independently to each induced subgraph.

Following the completion of the specified recursion depth, the accumulated separators define the global permutation $P$ for the sparse matrix. The downstream benefits are quantified by evaluating symbolic-factorization proxies, specifically the additional nonzeros generated through fill-in during $LDL^T$ factorization and the resulting operation counts. We also report end-to-end solver wall-clock times to provide a definitive comparison against the production-grade LS-GPart partitioner, assessing whether the quantum-enhanced partitions yield measurable improvements in solver efficiency.

\subsection{Generation of coarsened graphs}
\label{subsec:gen_graphs}

Industrially relevant finite element meshes often comprise $10^5$ to $10^8$ vertices, a scale that significantly exceeds the qubit capacities of current and near-term quantum hardware. Consequently, any viable quantum-classical workflow must incorporate a hardware-aware abstraction layer that maps high-dimensional problems onto constrained hardware topologies.

To address this, we employ a spectral coarsening strategy that contracts the original graph $G$ into a reduced surrogate graph $\tilde{G}$ while preserving its fundamental partitioning properties. In this framework, the quantum optimization is performed on the reduced instance, after which the resulting solution is projected—or "lifted"—back to the original high-resolution graph.

Beyond ensuring hardware compatibility, this reduction provides significant structural advantages. Graph order alone is an insufficient metric for partitioning complexity; many large-scale graphs contain weakly coupled or nearly separable sub-components whose bipartitions are computationally trivial. By coarsening the graph, these structurally simple regions are aggregated into supernodes, allowing the quantum resources to focus exclusively on the compact representation that captures the non-trivial global coupling.

The procedure initiates by embedding the vertex set into a low-dimensional Euclidean space using the smallest non-trivial eigenvectors of the graph Laplacian~\cite{godsil2001algebraic}. This construction is theoretically grounded in the continuous relaxation of the min-cut objective, where the optimal solutions are known to correspond to the low-frequency modes of the Laplacian. Consequently, these spectral coordinates encapsulate the dominant, large-scale topological structure of the graph. We subsequently apply $k$-means clustering to the spectral embedding, where $k$ is calibrated to the target coarse-graph size and the available qubit count. Each resulting cluster is contracted into a single supernode with a weight equal to the sum of its constituent vertex weights, and inter-cluster edges are aggregated to form the adjacency structure of $\tilde{G}$. Notably, this construction preserves the cut cost exactly: any bipartition of the coarse graph induces a corresponding bipartition of the original graph with an identical cut weight and balance ratio. To mitigate the sensitivity of $k$-means to stochastic initialization, we generate multiple candidate coarsenings from a fixed spectral embedding. These candidates are evaluated using a lightweight classical refinement to identify the most favorable proxy cut. Once a coarse bipartition is obtained—via either quantum or classical optimization—it is lifted to the original graph and undergoes final local improvement using the Fiduccia–Mattheyses (FM) heuristic. In Supplementary Information, we include a complete procedural logic of the spectral coarsening algorithm (\cref{alg:spectral-coarsening}).

\subsection{Application of Iterative-QAOA}

Iterative-QAOA is applied directly to the coarse graph instances produced by the spectral coarsening step described in \cref{subsec:gen_graphs}. The nodal and edge weights of the coarsened graphs are normalized by the maximum node and edge weight respectively. Each normalized coarse graph is then encoded as a cost Hamiltonian $H_C$ following the QUBO formulation of \cref{sec:background}, with the penalty coefficient $\lambda=1.0$ to enforce the nodal imbalance constraint. For instances up to 32 qubits, the quantum algorithm is executed using IonQ statevector simulators. Larger instances (up to 120-150 qubits) are handled by a matrix product state (MPS) simulator built on the QUIMB tensor-network library and NVIDIA's CUDA-Q/cuTensorNet library, using a fixed bond dimension of $\chi = 256$. All instances were executed on Nvidia A100 and H100 GPUs. Hardware validation on 36-qubit instances is performed on IonQ’s Forte trapped-ion QPU, described in \cref{sec:ionq-hardware}.

\begin{figure}[!htb]
    \centering
    \includegraphics[width=\columnwidth]{
    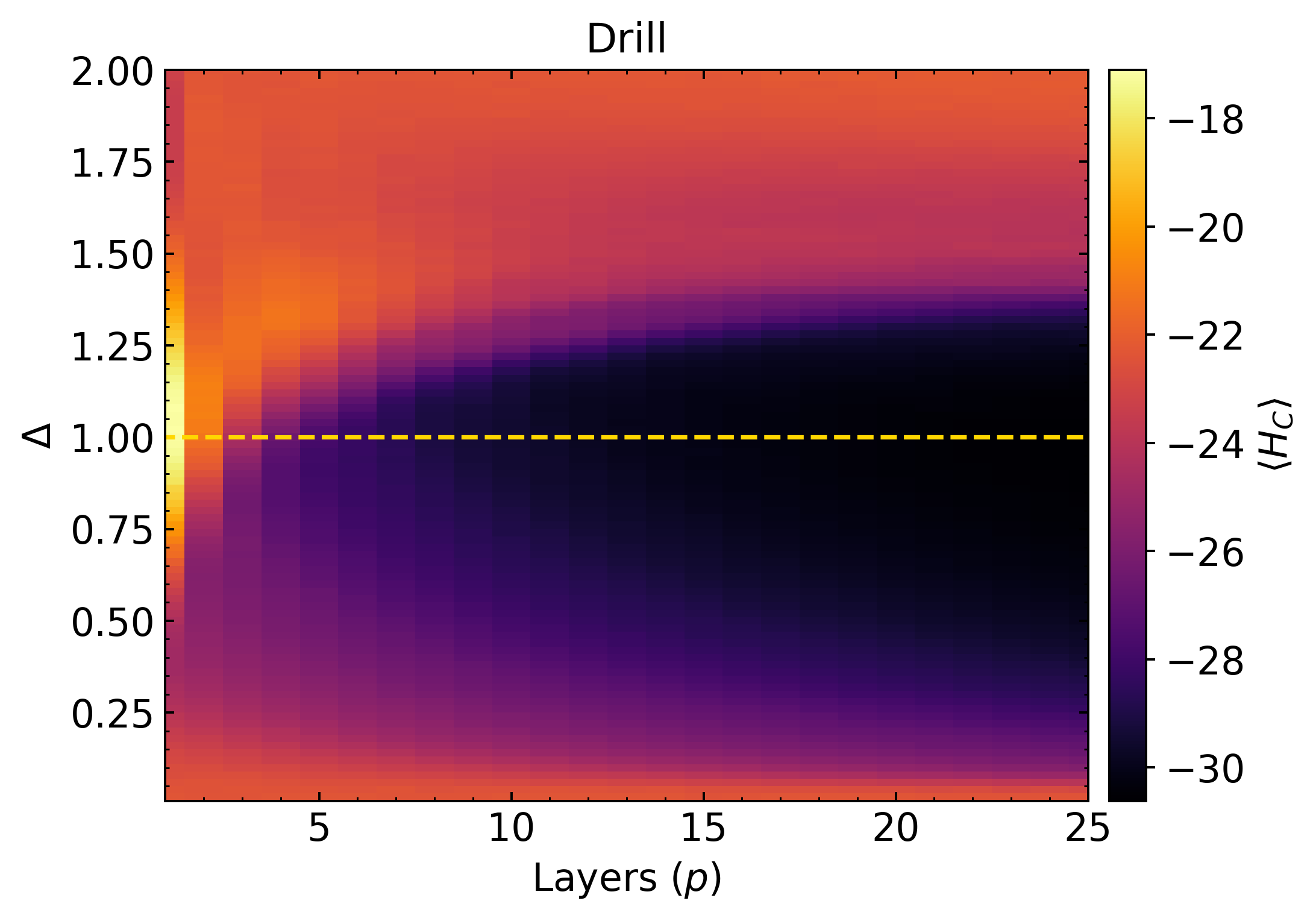}
    \caption{LR-QAOA performance landscape for a 24-qubit drill instance. The dashed horizontal line marks the optimal $\Delta = 1.0$ that minimizes $\ev{ H_C }$.}
    \label{fig:lrqaoa_param_exp}
\end{figure}

The ramp parameter $\Delta$ governing the linear-ramp angle schedule (\cref{eq:lr_schedule}) is selected via a parameter sweep described in \cref{sec:additional_results}. The sweep evaluates the cost Hamiltonian expectation value $\langle H_C \rangle$ as a function of $\Delta$ and circuit depth $p$ using noiseless statevector simulations on small instances for all four mesh models in \cref{tab:problem_instances}. For each model, the per-instance optimal $\Delta$ values obtained at 24, 28, and 32 qubits are then fit with a power-law model, which is used to extrapolate $\Delta$ to the larger graph sizes considered in the main experiments. \cref{fig:lrqaoa_param_exp} shows the optimal $\Delta$ parameter selection from the 24-qubit Drill problem instance. Plots for the other models can be found in \cref{fig:lrqaoa_param_exp_all} in \cref{sec:additional_results}.

The QAOA circuit is constructed with $p = 6$ layers for the JetEngine model and $p = 5$ layers for the rest using 5,000 measurement shots per circuit evaluation. The Hamiltonian block in the QAOA ansatz is constructed from a truncated number of terms from the problem Hamiltonian (which is expanded into a sum of Pauli terms) to control the number of 2 qubit gates used in the circuit. The number of terms used is set as $N_\mathrm{terms} = C\cdot N_\mathrm{qubits}$ where $N_\mathrm{qubits}$ is the number of qubits required to represent the problem with the constant $C$ set to 30 for the SedanCar model, 55 for the Drill and Impeller models and 80 for the JetEngine model. Truncation acts only on the two-qubit interactions, retaining the ZZ terms of largest coefficient magnitude, with $N_{\mathrm{terms}}$ the total two-qubit gate budget shared across the $p$ layers; the 36-qubit Drill and Impeller circuits thus contain 1,980 two-qubit gates and the 150-qubit JetEngine circuits 12,000. Sampled bitstrings are scored and refined against the full, untruncated objective, so truncation affects only the
sampling distribution, not the evaluation or selection of partitions.
After each circuit execution, the sampled bitstrings are passed through only one iteration of our custom Fiduccia–Mattheyses (FM) refinement algorithm to compensate for the fixed approximation (fixed bond dimension of $\chi = 256$) used in the MPS simulations. This is a light, greedy refinement that is local to the neighborhood of the sampled bitstring. The refined bitstrings are combined with the sampled bitstrings from the quantum circuit and sorted in increasing order of their energy. The first 50 lowest-energy states (bit strings) are identified and used to construct the Boltzmann-weighted warm-start state for the subsequent iteration, as described in \cref{sec:iterativeqaoa}. The Boltzmann weight $\beta_T$ is itself selected according to a schedule determined from the following equation: $\beta_T = 9x^2 + 1$ where $x$ is varied linearly from 0 to 1 in $N_\mathrm{iter}$ steps. This feedback loop is repeated for $N_\mathrm{iter}=10$ iterations, progressively concentrating the quantum state on high-quality partition candidates. 

\subsection{Classical heuristic refinement}
\label{subsec:post-process}

Partitions generated by the quantum solver generally necessitate additional refinement to ensure high-fidelity results. This requirement arises from two primary factors. First, solutions derived from the coarse graph and subsequently "lifted" to the original high-resolution graph are not inherently guaranteed to maintain local optimality at the full discretization level. Second, practical executions of Iterative-QAOA—whether performed on physical hardware or via approximate classical simulation—may yield partitions that, while near-optimal, are not strictly locally optimal with respect to single-vertex migrations. To address these limitations, we employ a customized variant of the Fiduccia–Mattheyses (FM) algorithm \cite{fiduccia1988linear} as a classical post-processing step. Our implementation adapts the standard FM framework to the specific demands of weighted industrial finite element graphs and the rigorous balance constraints required by Nested Dissection. This specific refinement procedure was previously validated in Ref.~\cite{aboumrad2025accelerating} and is utilized here to ensure methodological consistency.

As a linear-time local search heuristic, the FM algorithm iteratively migrates vertices between partitions to minimize the cut cost subject to a balance constraint. Our implementation diverges from the canonical formulation in two aspects to better suit industrial workloads. First, vertex moves are restricted to the partition with the higher aggregate weight. By always selecting from the heavier side, the algorithm actively steers toward balance throughout every pass, rather than allowing moves from either partition as in the standard formulation. Second, the algorithm enforces the balance constraint inline during best-gain tracking: a candidate move sequence is recorded as the new best only if the resulting partition satisfies the tolerance at that step. In contrast, the canonical FM algorithm tracks the best cumulative gain unconditionally and enforces balance post-hoc. Together, these two modifications ensure that feasibility is maintained throughout each pass rather than recovered after the fact, which improves stability for graphs with highly heterogeneous vertex weights. Within each pass, we retain the standard FM practices of cumulative gain tracking and prefix rollback, committing only move sequences that yield a net-positive improvement.
 
In the proposed hybrid workflow, this customized FM procedure is deployed in two distinct stages. The first is a coarse-scale screening applied to the reduced graph $\tilde{G}$ during the problem-space reduction phase to evaluate candidate coarsenings. The second is a full-scale refinement applied to the original graph $G$ following the lifting of the coarse solution. At this stage, the algorithm performs fine-grained structural adjustments that were inaccessible at the reduced resolution. We note that this study evaluates the full hybrid pipeline against LS-GPart and does not isolate the contribution of the coarse-graph solver; an ablation substituting a classical solver in the same slot, with all other stages fixed, is left to future work.

 \subsection{Defining the figures of merit}
\label{subsec:merit-factor-eval}

To assess the influence of the proposed partitioning strategies on downstream linear system performance, we utilized the LS-DYNA Nested Dissection workflow. This evaluation follows the integration methodology established in our previous work, where an external partitioner is interfaced directly with the LS-DYNA solver environment \cite{aboumrad2025accelerating}. Within this framework, LS-DYNA adopts the partitions generated by the Iterative-QAOA quantum algorithm to determine the matrix reordering. Subsequently, symbolic factorization is performed to calculate critical merit factors, specifically the increase in nonzeros due to fill-in and the total operation count required for the linear solve. Performance is further quantified by measuring the Factorization Wall-Clock Time (WCT) and the Total WCT for the complete Finite Element Analysis (FEA) simulation, allowing for a rigorous comparison against the internal LS-GPart heuristic under identical conditions. A distinct advantage of using the quantum algorithm is its ability to generate a set of high-quality candidate partitions, enabling the evaluation of merit factors across multiple solutions rather than relying on a single heuristic output. This process can be extended to level-$L$ Nested Dissection by executing LS-DYNA $L+1$ times. At each level $l$, the external partitioner is applied to the $2^{l-1}$ coarse graphs extracted during that stage; the resulting partitions are then returned to LS-DYNA to facilitate the generation of the next-level subgraphs until the final merit factors are obtained \cite{aboumrad2025accelerating}.

The total WCT of an FEA simulation is governed by a complex interplay of hardware and algorithmic variables. When executed across multiple compute nodes, performance is sensitive to the number of parallel processors, load balancing, inter-processor communication overhead, and network latency. Furthermore, WCT profiles vary significantly across different problem instances due to geometric complexity, structural symmetries, and total mesh size. For example, the SedanCar and JetEngine instances (Vibrational Analysis) differ in their computational footprint compared to the Impeller and Drill models (Transient Analysis). The proportion of WCT dedicated to matrix factorization is also problem-dependent. While factorization typically dominates the runtime for transient analysis problems, it accounts for approximately half of the total WCT in vibrational analysis cases. Although the heuristic steps within LS-DYNA between solving the Graph Partitioning Problem (GPP) and numerical factorization introduce an inherent uncertainty in the mapping of GPP solution quality to WCT, empirical evidence indicates a strong correlation. On average, partitions with lower graph cut costs and nodal weight imbalances consistently yield superior figures of merit. These observations suggest that leveraging quantum algorithms to find high-quality GPP solutions for larger-scale graphs will result in significantly improved performance and reduced simulation times.

 \subsection{IonQ quantum hardware}\label{sec:ionq-hardware}

Experimental data for this study were obtained using IonQ’s Forte quantum processing unit (QPU)~\cite{Chen2024-co}, a system featuring 36 qubits encoded in the hyperfine ground states of trapped $^{171}\text{Yb}^+$ ions. These ions are generated via laser ablation and selective ionization and are confined within compact, integrated vacuum packages utilizing surface linear Paul traps. Universal control is achieved through two-photon Raman transitions driven by 355 nm laser pulses, which facilitate a native gate set comprised of arbitrary single-qubit rotations and entangling $ZZ$ gates. To ensure precision and scalability, the QPU integrates an advanced optical control architecture based on acousto-optic deflectors (AODs) that enable independent beam steering to individual ions. This configuration significantly reduces beam alignment errors~\cite{Kim:2008ApOpt, Pogorelov:2021PRXQ} and is supported by automated calibration software to maintain stable operations. The IonQ Forte system showed sustained high-fidelity performance, with median direct randomized benchmarking fidelities for two-qubit gates of $99.3\%$ at typical gate durations of $950~\mu\text{s}$, while single-qubit gates achieve fidelities of approximately $99.98\%$ with operation times of $130~\mu\text{s}$. To further refine the output, data processing includes a sample debiasing protocol in which the total shot count is partitioned into batches of 25 and averaged across these groups~\cite{maksymov2023}. 
Each batch is executed under an independently symmetrized realization of the circuit, so averaging across batches suppresses coherent, qubit-specific systematic errors. No further error mitigation was applied: Iterative-QAOA requires only that low-energy bitstrings appear in the sample, not accurate expectation values. And residual local defects are repaired by the downstream FM refinement. A quantitative comparison of
mitigation strategies is beyond the scope of this study.
This robust architecture supports the consistent high gate fidelities necessary for complex quantum optimization experiments characterized by dense interaction structures, providing a stable platform for the computational tasks executed in this work.

\section{Results}\label{sec:results}

\subsection{Results from simulations and hardware execution}\label{sec:noiseless_sims}

\begin{figure*}[!htb]
    \centering
    \includegraphics[width=\textwidth]{
    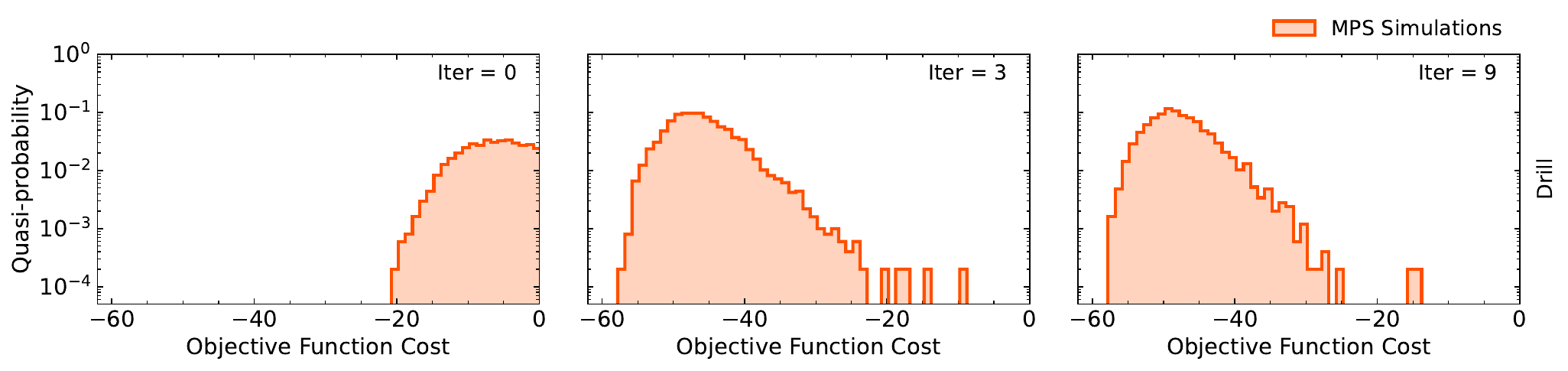}
    \caption{Iterative-QAOA executed on a 120-qubit Drill problem instance using the NVIDIA CUDA-Q/cuTensorNet MPS simulator with a bond dimension $\chi = 256$. Each panel shows the cost probability distributions at the initial (Iter = 0), an intermediate (Iter = 3), and the final (Iter = 9) iteration. The number of layers in the QAOA circuit was $p = 5$. The algorithm parameters used are $\Delta = 0.3$.}
    \label{fig:energy_hist_drill}
\end{figure*}

\begin{figure*}[!htb]
    \centering
    \includegraphics[width=\textwidth]{
    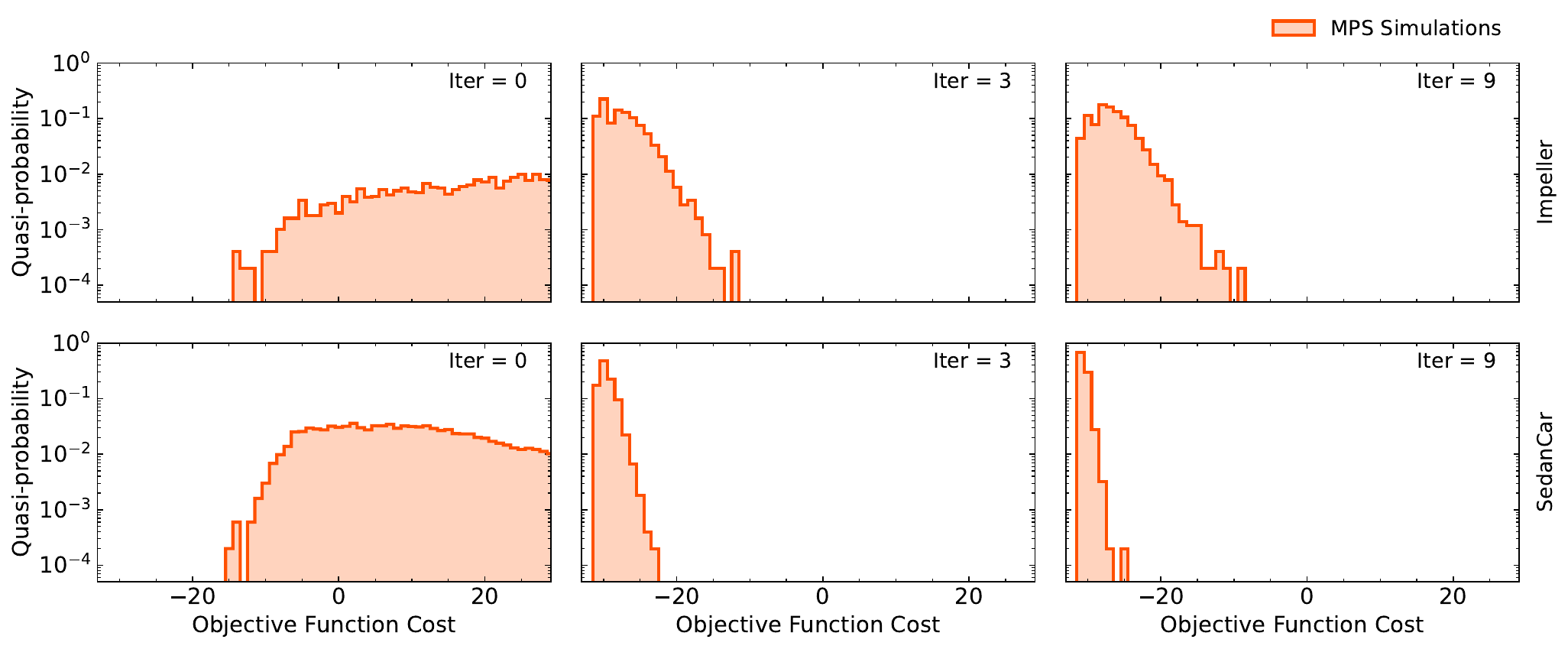}
    \caption{Iterative-QAOA executed on a 120-qubit Impeller and SedanCar problem instances using the NVIDIA CUDA-Q/cuTensorNet MPS simulator with a bond dimension $\chi = 256$. Each panel shows the cost probability distributions at the initial (Iter = 0), an intermediate (Iter = 3), and the final (Iter = 9) iteration. The number of layers in the QAOA circuit was $p = 5$. The algorithm parameters used are $\Delta = \{1.0, 0.64\}$ for the Impeller and SedanCar models respectively.}
    \label{fig:energy_hist_honda-impeller}
\end{figure*}

To investigate the performance across varying problem scales, coarsened graphs ranging from 24 to 150 nodes were generated for each of the four problem instances. The Iterative-QAOA algorithm was subsequently applied to these coarsened partitioning problems, encompassing 63 distinct simulations. The 36 node problem instances for the Drill, Impeller and SedanCar models were also executed on IonQ Forte quantum hardware. 
Hardware validation is limited to 36-node instances because each graph
node is encoded in one qubit and Forte provides 36 qubits; all results at
larger node counts are obtained from noiseless classical simulation
(statevector up to 32 qubits, MPS beyond) and are labeled accordingly in
Figs.\ref{fig:total_wct_reduction} and \ref{fig:fact_wct_reduction}. 
While the protocol was configured for $N_{\mathrm{iter}} = 10$ iterations, the low-energy solutions for both the classical simulations and hardware executions consistently reached convergence in fewer than five iterations. \cref{fig:energy_hist_drill} illustrates these results for a 120-node instance of the Drill problem, providing three sequential plots that trace the sampled bitstring distributions—both raw and refined—from the initial, intermediate, and final iterations. Comparable results for the remaining problem instances are detailed in \cref{fig:energy_hist_honda-impeller} within \cref{sec:additional_results}.

As evidenced by the data in \cref{fig:energy_hist_drill,fig:energy_hist_honda-impeller}, the algorithm effectively "squeezes" the probability distribution toward solutions with lower graph partition costs, resulting in a final state highly skewed toward the low-energy sector. This iterative shift is facilitated by the biased warm-starting of the QAOA circuit, which progressively refines the distribution until the low-energy sector converges and no further novel solutions are produced. With subsequent iterations, the probability of sampling these high-quality solutions continues to rise, causing the entire distribution to peak sharply within the low-energy regime.

Despite this success, a degradation in solution quality will arise as the graph size increases to around 100 nodes and beyond when using the Matrix Product State (MPS) simulator (for more details, see Ref.~\cite{lopez2025non}). One of the reasons is that QAOA ansatz depths of $p=5$ and $p=6$ layers may lack the expressivity required to resolve the ground state for problems at this scale and will likely require more layers. However, increasing circuit depth generates greater entanglement, which necessitates a larger bond dimension to maintain the fidelity of the MPS approximation, thereby increasing the demand for classical computational resources. As discussed in detail by Ref.~\cite{lopez2025non}, these findings suggest that at a fixed bond dimension, MPS simulations may not fully capture the complex correlations introduced by deeper circuits. This limitation underscores the critical requirement for large-scale quantum hardware characterized by high gate fidelities and robust error mitigation or error correction. 
Instances within reach of current hardware were executed on hardware,
and larger instances through classical simulation. Because MPS at fixed
bond dimension $\chi=256$ is classically tractable by construction, the
120--150-qubit results characterize the algorithm's scaling rather than
demonstrate a quantum capability. The prospective value of the workflow
lies at scales beyond classical simulability, where the observed growth of
gains with node count, if it persists, could be realized only on quantum
hardware.
Nevertheless, the ability of Iterative-QAOA to consistently yield high-quality solutions under these constrained conditions serves as a promising indicator of its practical applicability to large-scale optimization tasks.

\subsection{Evaluation of figures of merit}
\label{sec:merit-figures}

\begin{figure*}[!htb]
    \centering
    \includegraphics[width=\textwidth]{
    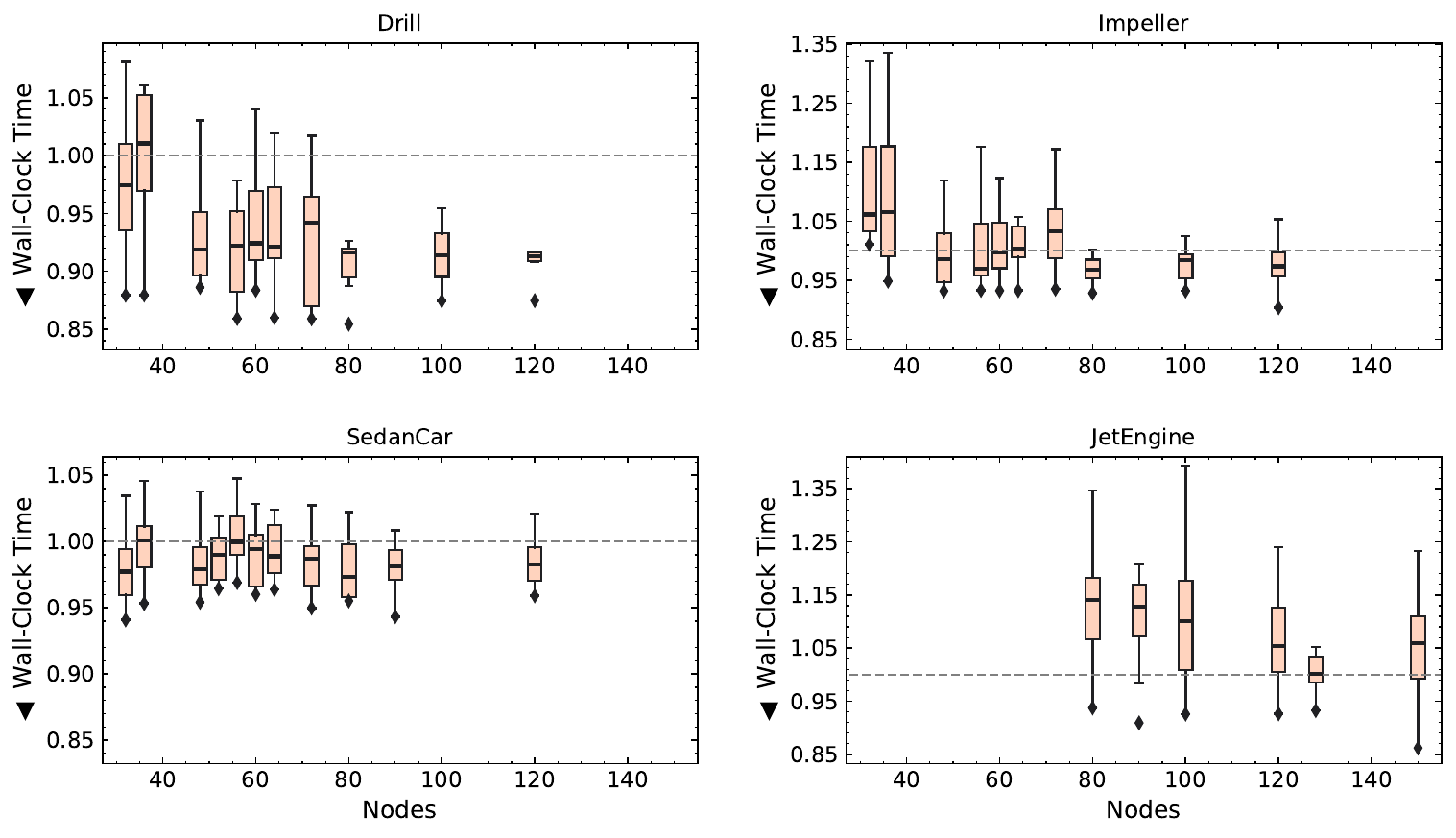}
    \caption{Total wall-clock time (WCT) reduction as a function of coarse graph size using quantum-derived partitions. The horizontal dashed line denotes the normalized baseline ($1.0$), established by the optimal total WCT achieved via the internal LS-DYNA partitioner on the original $10,000$-node baseline graph. All relative performance data are plotted with respect to this baseline. For each graph size, box plots represent the distribution of the total WCT outcomes (with whiskers showing the 90th and 10th percentiles, box boundaries showing the 75th and 25th percentiles, and line showing the median) from 20 highest-quality partitions, with diamond markers highlighting the absolute minimum total WCT attained. Although inherent stochasticity is observed due to the factors discussed in \cref{subsec:merit-factor-eval}, the cumulative data indicate a discernible trend toward improved total WCT as the coarsened graph dimensionality increases.}
    \label{fig:wct_boxplots}
\end{figure*}

\begin{figure}[!htb]
    \centering
    \includegraphics[width=\columnwidth]{
    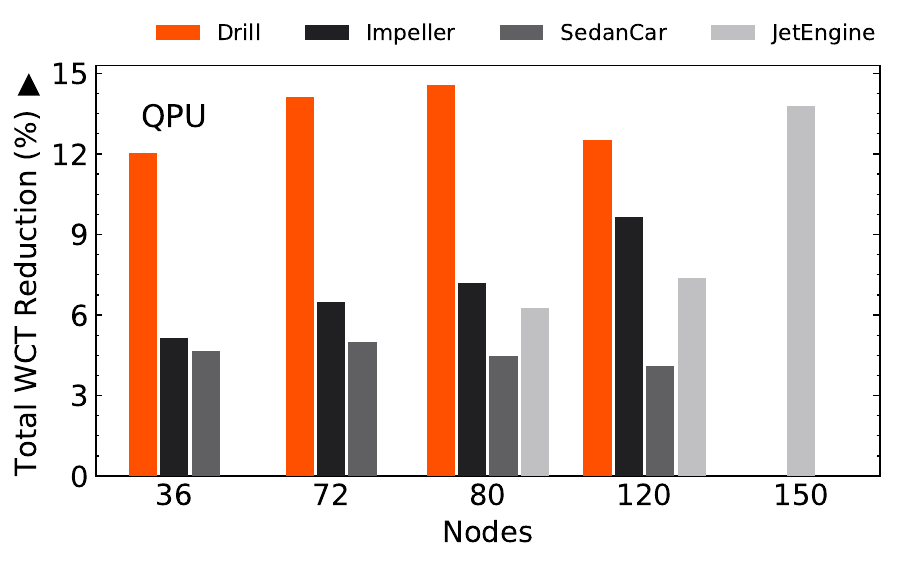}
    \caption{Maximum percentage reduction in total wall-clock time (WCT) across the different models and coarse graph sizes. Results illustrate performance gains achieved using partitions generated by the quantum algorithm executed on IonQ Forte QPU for 36-node graphs, and NVIDIA's CUDA-Q/cuTensorNet Matrix Product States (MPS) simulator on NVIDIA A100/H100 GPUs for larger graphs. Improvement is expressed as a percentage relative to the baseline, which is defined as the minimum total WCT attained using the optimal internal LS-DYNA partition of the original $10,000$-node graph. }
    \label{fig:total_wct_reduction}
\end{figure}

To quantify the performance of the proposed methodology, quantum-derived partitions from each coarsened graph instance were subjected to the lifting and refinement procedure detailed in \cref{subsec:post-process}. For each problem instance, the 20 highest-quality partitions were selected to calculate the figures of merit (\cref{subsec:merit-factor-eval}). Finite Element Analysis (FEA) was conducted using LS-DYNA in distributed-memory mode via Message Passing Interface (MPI). The SedanCar model was executed on an AWS compute cluster utilizing a single MPI rank. For the vibrational analysis models, 20 eigenmodes were computed to minimize evaluation time; while production vibrational analysis typically may require hundreds of modes, the total wall-clock time (WCT) in this regime increases fairly linearly with the number of modes.

Due to the proprietary nature of the remaining models, the Drill, Impeller, and JetEngine were evaluated on Synopsys infrastructure. The cluster environment consisted of nodes equipped with two 16-core Intel Xeon Gold 6142 CPUs and 512GB of RAM. We employed 16 MPI ranks for the Impeller and Drill models and 64 ranks for the JetEngine. To manage computational overhead, transient simulations were limited to 2 timesteps, whereas typical production runs might simulate hundreds of timesteps over several days. To maximize the impact of the quantum partitions while minimizing QPU resources, the quantum algorithm was applied only to the first level of the Nested Dissection algorithm; all subsequent levels were resolved using the internal LS-GPart solver.

\cref{fig:wct_boxplots} illustrates the normalized total WCT of the FEA simulations using refined quantum partitions as a function of the coarsened graph node count. The horizontal dashed line (normalized to 1.0) represents the baseline WCT achieved by applying the internal LS-GPart partitioner entirely to the original 10,000-node graph. Box plots delineate the distribution of the 20 best partitions—representing the $10^{\text{th}}$ and $90^{\text{th}}$ percentiles—while diamond markers indicate the absolute minimum WCT attained. The maximum reduction in WCT achieved as a function of the number of nodes in the coarsened graphs is shown in \cref{fig:total_wct_reduction}. In hardware executions on the IonQ Forte QPU utilizing 36-node coarse graphs, we observed a WCT reduction of about 12\% for the Drill model, with more modest improvements for the Impeller and SedanCar. Further benchmarks using the NVIDIA CUDA-Q/cuTensorNet MPS simulator yielded a maximum WCT reduction of about 14.6\% for the Drill model, 13.8\% for the JetEngine model, 9.7\% for the Impeller model, and 5.9\% for the SedanCar model. A detailed table of wall-clock times for all instances is shown in Table~\ref{tab:wct_all}. 
These headline figures are best-case values: the single best of the 20 highest-quality partitions per configuration (diamond markers in Fig. \ref{fig:wct_boxplots}). The improvement is not uniform: at the smallest coarse-graph sizes (32–36 nodes) the median partitions offer little or no gain over the LS-GPart baseline, and for the JetEngine model the gains are largely confined to the best candidates. The attainable reduction grows with coarse-graph node count, i.e., with the number of qubits employed (Figs. \ref{fig:wct_boxplots} and \ref{fig:total_wct_reduction}).

To put these percentages in production terms: the measured runs use
reduced workloads (2 transient timesteps; 20 eigenmodes), but the reduction
applies to the dominant, repeated solve phase, so at production scale---where
a large transient analysis of the Drill model can require roughly seven days
of compute---the 14.6\% best-case reduction would correspond to about one day
of saved execution time. All comparisons are against the production LS-GPart
partitioner operating on the same $10{,}000$-node baseline graph under
identical LS-DYNA configuration, MPI layout, and compute hardware.

Interestingly, a direct 1:1 correspondence between symbolic factorization merit figures (such as ``fill-in'' non-zeros and number of operations) and actual WCT was only observed for the single MPI rank SedanCar model execution. For the other simulations utilizing multiple MPI ranks, this relationship is complicated by parallel execution effects such as load balancing and network latency. The identification of a predictive mapping between merit figures and parallel runtime is reserved for future study of LS-DYNA's internal mechanisms. However, even with the inherent stochasticity discussed in \cref{subsec:merit-factor-eval}, the cumulative data demonstrate a discernible trend: total WCT improves as the number of nodes of the coarsened graph increases. This provides strong evidence that further gains may be realized as higher-quality quantum hardware enables the processing of even larger coarsened graphs.

\begin{figure}[!htb]
    \centering
    \includegraphics[width=\columnwidth]{
    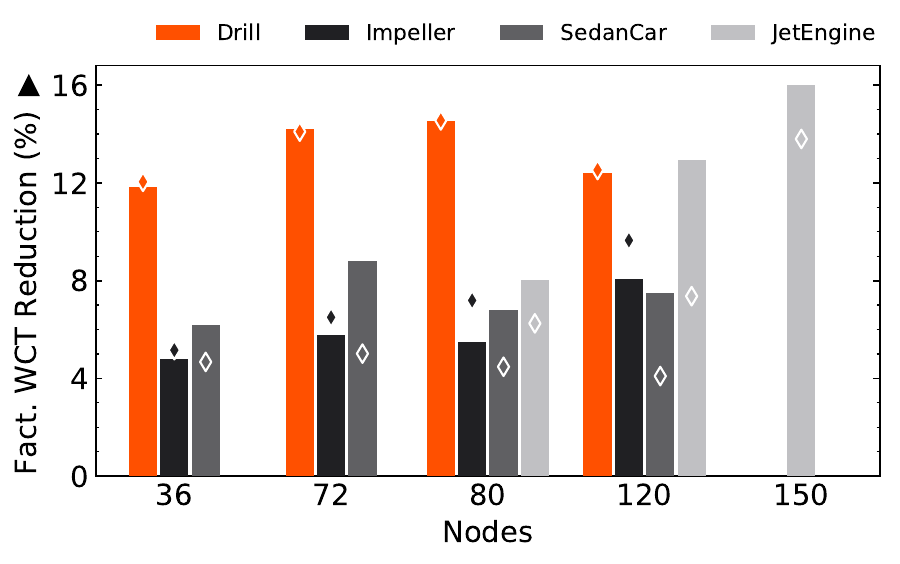}
    \caption{Maximum percentage reduction in factorization wall-clock time (WCT) across the different models and coarse graph sizes. Bar heights are measured relative to the factorization-WCT baseline, defined as the minimum factorization WCT attained using the optimal internal LS-DYNA partition of the original $10,000$-node graph. For comparison, diamond markers reproduce the total-WCT reduction percentages from \cref{fig:total_wct_reduction}; these markers are referenced to the total-WCT baseline and are overlaid only to compare total and factorization improvements. Results include IonQ Forte QPU executions for the 36-node cases and NVIDIA's CUDA-Q/cuTensorNet Matrix Product States (MPS) simulator on NVIDIA A100/H100 GPUs for the larger graphs.}
    \label{fig:fact_wct_reduction}
\end{figure}

\begin{figure*}[!htb]
    \centering
    \includegraphics[width=\textwidth]{
    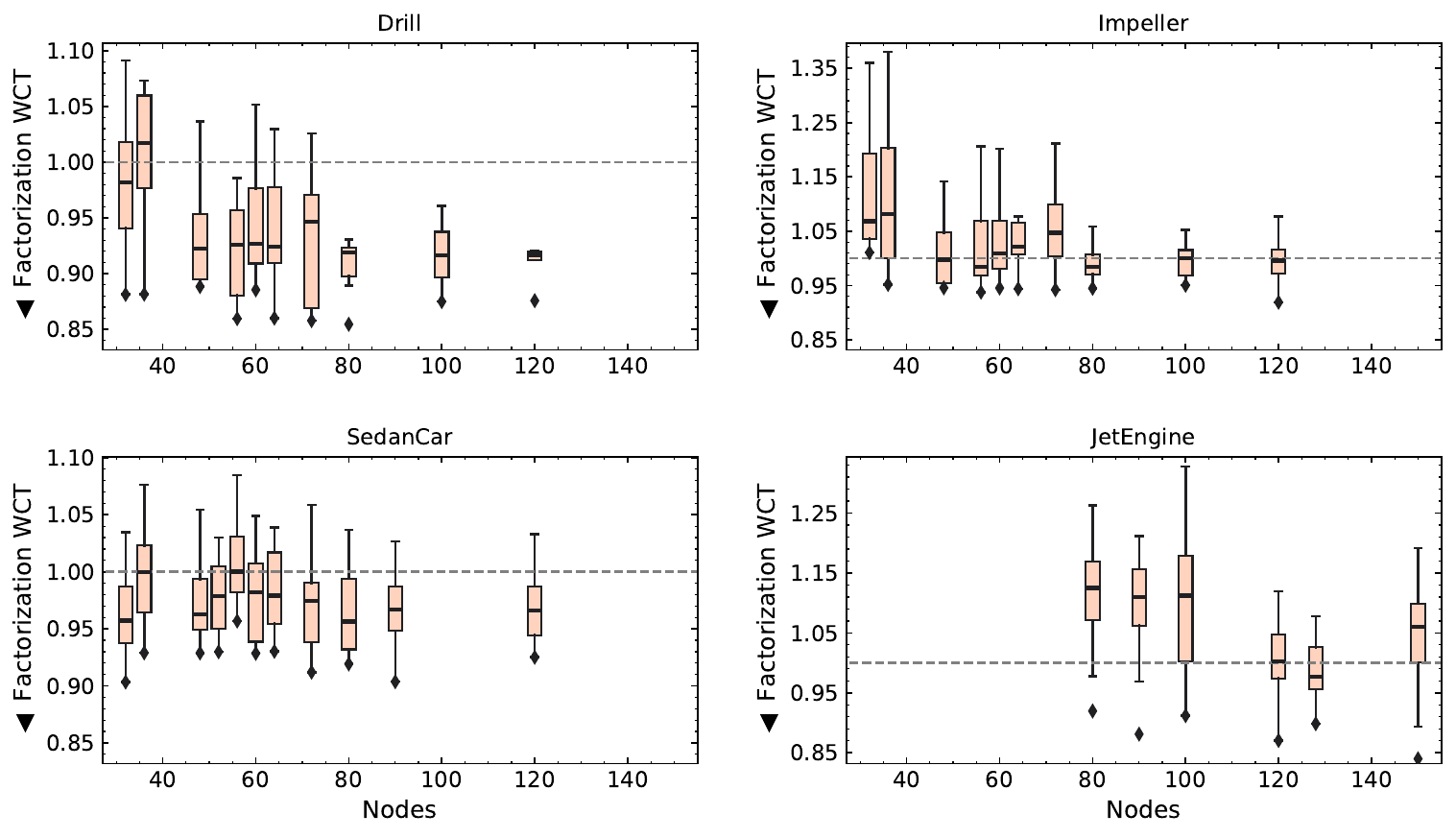}
    \caption{Factorization wall-clock time (WCT) reduction as a function of coarse graph size using quantum-derived partitions. The horizontal dashed line denotes the normalized baseline ($1.0$), established by the optimal factorization WCT achieved via the internal LS-DYNA partitioner on the original $10,000$-node baseline graph. All relative performance data are plotted with respect to this baseline. For each graph size, box plots represent the distribution of the factorization WCT outcomes (specifically the 10th and 90th percentiles) from the 20 highest-quality partitions, with diamond markers highlighting the absolute minimum factorization WCT attained. Although inherent stochasticity is observed due to the factors discussed in \cref{subsec:merit-factor-eval}, the cumulative data indicate a discernible trend toward improved factorization WCT as the coarsened graph dimensionality increases.}
    \label{fig:fact_wct_boxplots}
\end{figure*}

These results highlight a significant distinction in problem suitability. In vibrational analysis models, factorization WCT accounts for only about half of the total runtime, whereas in transient analysis models, it constitutes nearly the entire WCT (see \cref{fig:fact_wct_reduction,fig:fact_wct_boxplots}). Consequently, the substantial gains in factorization efficiency achieved for vibrational models do not translate as effectively to total WCT as they do for transient problems. This suggests that the transient analysis class—and other regimes dominated by factorization time—may be particularly well-suited for quantum enhancement. Furthermore, the superior reductions observed in the Drill and JetEngine models may stem from their inherent geometric asymmetry. While classical heuristics are highly effective on more symmetric structures like the Impeller and SedanCar, they often struggle with complex, asymmetric geometries, providing a clear opportunity for quantum optimization to deliver superior partitioning results and larger reductions in the total FEA execution time.

\section{Conclusions and Outlook}\label{sec:future-outlook}

In this study, we have demonstrated the efficacy of the Iterative-QAOA quantum algorithm in solving the graph partitioning problems that underpin large, sparse matrix solutions in FEA solvers like LS-DYNA. By utilizing a non-variational approach, we have established a highly scalable methodology that remains resilient against the common limitations of traditional variational quantum algorithms, such as barren plateaus in classical parameter optimization, the need for excessive circuit executions, and the inherent uncertainty in quantum ansatz definition. By integrating this algorithm into a production-level, hybrid quantum-classical workflow, we have shown that quantum acceleration is a viable path for accelerating the solution of complex linear systems across diverse domains.

Our experimental results, which utilized both IonQ Forte QPUs and Matrix Product State (MPS) simulations on NVIDIA A100/H100 GPUs via the NVIDIA CUDA-Q/cuTensorNet libraries, pushed the boundaries of our integrated quantum-classical pipeline to 150 qubits. This rigorous testing showed that Iterative-QAOA consistently identifies high-quality separators for the Nested Dissection algorithm, directly resulting in up to a 14.6\% improvement in total wall-clock time (WCT) for the Drill model and 13.8\% for the JetEngine model relative to the classical heuristic within LS-DYNA. Despite the introduction of noise at large graph sizes due to fixed bond dimensions in MPS simulations, we observed a persistent trend: as the number of graph nodes increases, so does the efficiency of the FEA simulation. This provides strong evidence that further performance gains are possible as quantum hardware continues to scale in fidelity and capacity. 
While these results do not constitute a claim of quantum advantage---and
further classical optimization at the coarse level could narrow the observed
gains---the measured benefit and its systematic growth motivate evaluation
on other quantum processors at larger qubit scales.

Crucially, this research has identified specific classes of FEA problems, such as vibrational and transient analysis, that are exceptionally amenable to quantum enhancement. By offloading a single graph partitioning step to the quantum computer during the reordering process, the resulting matrix configuration can be utilized across many subsequent classical FEA solver iterations. This allows the quantum execution time to be effectively amortized over the entire FEA simulation. For example, in a large-scale transient evolution analysis of a Drill model, which can require seven days of compute, our 14.6\% improvement translates to a full day of saved execution time. Furthermore, our data indicates that complex, non-symmetric geometries, where classical heuristics often struggle, offer the greatest potential for quantum-driven reordering to provide a superior impact on total WCT.

Looking ahead, future work will focus on investigating a broader array of industrial use cases across different domains to further isolate the problem classes most receptive to quantum acceleration. Leading solvers such as Ansys Fluent, HFSS, CFX, COMSOL Multiphysics, and Dassault’s Abaqus all face the fundamental challenge of processing massive physical models with extreme speed and accuracy. Since these platforms rely on the solution of large, sparse linear systems on large scale compute clusters, integrating our quantum-classical approach into this pipeline represents a promising path toward delivering superior performance and  substantial gains not only in time-to-solution but also in reduced execution costs and lower energy consumption for large-scale industrial simulations.
Full reproduction of this study requires the modified
LS-DYNA build, three proprietary benchmark models used with permission, and
access to IonQ Forte hardware. The algorithm and its hyperparameters are,
however, fully specified in this paper, the SedanCar model is based on a
publicly available crash-simulation mesh, and the simulations rely only on
open-source libraries.

\section*{Acknowledgments}

We thank Richard Sturt (ARUP) for providing us with the model based on the NHTSA Honda Accord model, James Ong (Rolls-Royce) for providing us with the Jet Engine and Impeller models, and George Laird (Predictive Engineering) for providing us with the Drill model. We thank Michael Brett, Eric Kessler, Crystal Mascareno, and Joseph Smalzer for enabling performance evaluation of the end-to-end LS-DYNA workflow on an AWS cluster. We thank Dmitri Liakh for helping us optimize environment variables for large scale simulations with NVIDIA cuTensorNet.

\section{Supplementary Information}\label{SI}

\begin{algorithm}[!htb]
\caption{Spectral Graph Coarsening with Screening}
\label{alg:spectral-coarsening}
\begin{algorithmic}[1]
\Require Graph $G=(V,E)$; target coarse size $k$; spectral dimension $d$;
         screening rounds $N_{\mathrm{screen}}$
\Ensure  Best assignment $\sigma^*:V\!\to\!\{1,\dots,k\}$; coarse graph $\tilde{G}^*$

\State Compute Laplacian $L$.
\State Use Lanczos algorithm to compute the $d$ nontrivial smallest eigenvectors.
\State Form spectral embedding $\Phi\in\mathbb{R}^{|V|\times d}$ from these eigenvectors.
\State $c^* \gets \infty$; \quad $\sigma^* \gets \varnothing$; \quad $\tilde{G}^* \gets \varnothing$
\For{$t=1,\dots,N_{\mathrm{screen}}$}
    \State $\sigma^{(t)} \gets k$-means$(\Phi)$ \Comment{random seed}
    \State Construct coarse graph $\tilde{G}^{(t)}$ by aggregating node weights within clusters and summing edge weights between clusters.
    \State Generate random balanced partitions $\mathcal{P}$ of $\tilde{G}^{(t)}$. \Comment{balanced cardinality}
    \State Refine $\mathcal{P}$ using modified-FM. \Comment{see main text}
    \State $c^{(t)} \gets$ best cost among refined partitions in $\mathcal{P}$.
    \If{$c^{(t)} < c^*$}
        \State $c^* \gets c^{(t)}$; $\sigma^* \gets \sigma^{(t)}$; $\tilde{G}^* \gets \tilde{G}^{(t)}$
    \EndIf
\EndFor
\State \Return $\sigma^*,\tilde{G}^*$
\end{algorithmic}
\end{algorithm}

The procedural logic of the proposed coarsening approach is detailed in \cref{alg:spectral-coarsening}. To assess the performance of the proposed spectral coarsening method, we benchmarked the quality of the resulting coarse graphs against reference graphs generated by LS-DYNA, utilizing the Drill mesh as a primary test case. The primary metric for this evaluation is the graph partition cost. Starting from a baseline graph of 10,000 nodes—a common coarse graph size within the LS-DYNA framework—we generated a suite of reduced graphs ranging from 24 to 500 nodes using both our spectral approach and the standard LS-DYNA coarsening utility.

For each instance produced via spectral coarsening, we refined 1,000 cardinality-balanced random initial partitions using our modified Fiduccia-Mattheyses (FM) algorithm (see \cref{subsec:post-process}). The performance of our method was then evaluated by comparing the minimum partition costs achieved for these refined spectral instances against the partition costs obtained from the corresponding LS-DYNA coarse graphs partitioned using the internal LS-GPart algorithm. This comparative framework allows for a direct assessment of how effectively each coarsening strategy preserves the connectivity features of the original mesh at significantly reduced scales. Our graph coarsening method showed improved performance over the reference coarsened graphs from LS-DYNA for most of the coarsened graph sizes.

\subsection{Additional results}\label{sec:additional_results}

\begin{table}[!ht]
    \centering
    \caption{Optimal LR-QAOA ramp parameter $\Delta$ for each model and small instance size explored.}
    \begin{tabular*}{\columnwidth}{@{\extracolsep{\fill}}lrrrr}
        \toprule
        \multirow{2}{*}{Nodes} & \multicolumn{2}{c}{Vibrational Analysis} & \multicolumn{2}{c}{Transient Analysis} \\ 
        \cmidrule(lr){2-3} \cmidrule(lr){4-5}
        & SedanCar & JetEngine & Impeller & Drill \\ 
        \midrule
        24 & 1.30 & 1.08 & 1.14 & 1.00 \\
        28 & 1.20 & 1.14 & 1.10 & 0.86 \\
        32 & 1.15 & 1.26 & 1.08 & 0.90 \\
        \bottomrule
    \end{tabular*}
\label{tab:lrqaoa_deltas}
\end{table}

\begin{figure*}[h!]
    \centering
    \includegraphics[width=\textwidth]{
    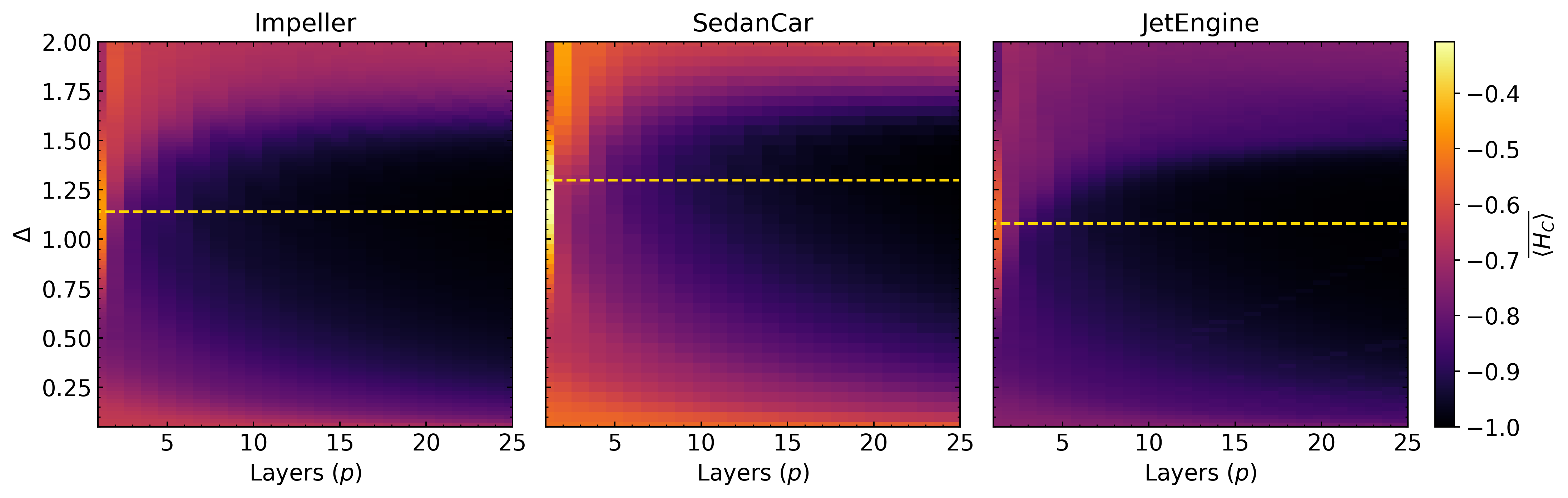}
    \caption{LR-QAOA cost landscape for 24-qubit instances of the Impeller, SedanCar, and JetEngine models, as a function of the ramp parameter $\Delta$ and circuit depth $p$. The color scale of each panel is normalized as $\overline{\langle H_C\rangle} \equiv \langle H_C\rangle/\max\langle H_C\rangle$. Results are from noiseless statevector simulations. All three instances exhibit a coherent low-energy valley whose precise minima are located at $\Delta=1.14$, $1.30$, and $1.08$, from left to right respectively.}
    \label{fig:lrqaoa_param_exp_all}
\end{figure*}

In the LR-QAOA schedule of \cref{eq:lr_schedule}, a single scalar $\Delta$ simultaneously sets the entire $(\boldsymbol{\gamma}, \boldsymbol{\beta})$ parameter set; choosing the optimal $\Delta$ is therefore the central configuration task. To do so, we sweep $\Delta$ jointly with the circuit depth $p$ and compute the expectation value $\langle H_C \rangle$ with noiseless statevector simulations. For each of the four mesh models (SedanCar, JetEngine, Impeller, and Drill), this grid search is repeated on instances of 24, 28, and 32 qubits; the per-instance optimal $\Delta$ is taken as the minimizer of $\langle H_C \rangle$. The values of these parameters are shown in \cref{tab:lrqaoa_deltas}. A representative set of parameter landscapes is shown in \cref{fig:lrqaoa_param_exp_all} for the 24-qubit instances of the Impeller, SedanCar, and JetEngine models. Across all instances, the landscapes exhibit a consistent ``performance valley'', a region of low $\langle H_C \rangle$ that persists as $p$ increases, whose precise minimum varies slightly between models. By utilizing these parameters at the smaller node sizes we can fit a power law model which can then be used to extrapolate the parameters to any graph size. This ensures a high-fidelity mapping of the LR-QAOA schedule providing a robust foundation for the larger-scale executions.

\begin{table*}[!htb]
\caption{Minimum total and factorization wall-clock times (WCT) for Vibrational Analysis and Transient Analysis simulations. All times are reported in seconds (s). End-to-end simulations of SedanCar were performed on an AWS cluster and end-to-end simulations of the other 3 models were performed on a Synopsys cluster as described in the text. Entries with "--" indicate experiments were not performed for that configuration. For vibrational simulations, 20 eigenmodes were computed, while production vibrational analysis typically require hundreds of modes. For transient simulations, 2 timesteps were computed, while production transient analysis typically requires hundreds of timesteps over several days. Across the configurations reported here, the best total-WCT reduction is 14.6\%, while the best 36-node hardware result is approximately 12\%.}
\centering
    \begin{tabular*}{\textwidth}{@{\extracolsep{\fill}}l rr rr rr rr}
    \toprule
    & \multicolumn{4}{c}{Vibrational Analysis} & \multicolumn{4}{c}{Transient Analysis} \\
    \cmidrule(lr){2-5} \cmidrule(lr){6-9} \\[-1em]
    & \multicolumn{2}{c}{SedanCar} & \multicolumn{2}{c}{JetEngine} & \multicolumn{2}{c}{Impeller} & \multicolumn{2}{c}{Drill} \\
    \cmidrule(lr){2-3} \cmidrule(lr){4-5} \cmidrule(lr){6-7} \cmidrule(lr){8-9} \\[-1em]
    Nodes & \makecell[r]{Total \\ WCT [s]} & \makecell[r]{Factorization \\ WCT [s]} & \makecell[r]{Total \\ WCT [s]} & \makecell[r]{Factorization \\ WCT [s]} & \makecell[r]{Total \\ WCT [s]} & \makecell[r]{Factorization \\ WCT [s]} & \makecell[r]{Total \\ WCT [s]} & \makecell[r]{Factorization \\ WCT [s]}  \\
    \midrule
    32 & 10,828.30 & 5,552.73 & -- & -- & 1,664.56 & 1,469.09 & 1,863.54 & 1,759.05 \\
    36 & 10,970.10 & 5,709.86 & -- & -- & 1,561.92 & 1,383.90 & 1,863.88 & 1,759.44 \\
    48 & 10,981.20 & 5,709.05 & -- & -- & 1,534.30 & 1,374.77 & 1,877.89 & 1,773.16 \\
    52 & 11,099.00 & 5,715.59 & -- & -- & -- & -- & -- & -- \\
    56 & 11,152.10 & 5,881.37 & -- & -- & 1,536.53 & 1,363.22 & 1,820.93 & 1,715.74 \\
    60 & 11,049.50 & 5,707.16 & -- & -- & 1,535.25 & 1,374.21 & 1,872.57 & 1,767.30 \\
    64 & 11,090.80 & 5,717.91 & -- & -- & 1,535.67 & 1,372.23 & 1,822.16 & 1,716.63 \\
    72 & 10,931.10 & 5,605.63 & -- & -- & 1,539.77 & 1,369.82 & 1,820.34 & 1,712.14 \\
    80 & 10,992.90 & 5,650.08 & 6,450.05 & 1,985.40 & 1,528.40 & 1,373.56 & 1,810.63 & 1,705.60 \\
    90 & 10,856.00 & 5,554.54 & 6,255.39 & 3,586.24 & -- & -- & -- & -- \\
    100 & -- & -- & 6,370.31 & 3,696.84 & 1,534.53 & 1,381.72 & 1,853.12 & 1,746.23 \\
    120 & 11,036.90 & 5,686.27 & 6,373.53 & 3,542.98 & 1,487.96 & 1,336.17 & 1,853.64 & 1,747.90 \\
    128 & -- & -- & 6,415.44 & 3,657.03 & -- & -- & -- & -- \\
    150 & -- & -- & 5,930.74 & 3,418.56 & -- & -- & -- & -- \\
    \midrule \\[-1em]
    LS-GPart & 11,508.20 & 6,146.57 & 6,880.27 & 4,069.80 & 1,646.86 & 1,453.59 & 2,119.19 & 1,995.82 \\
    \bottomrule
    \end{tabular*}
    \label{tab:wct_all}
\end{table*}

\begin{figure*}[!htb]
    \centering
    \includegraphics[width=\textwidth]{
    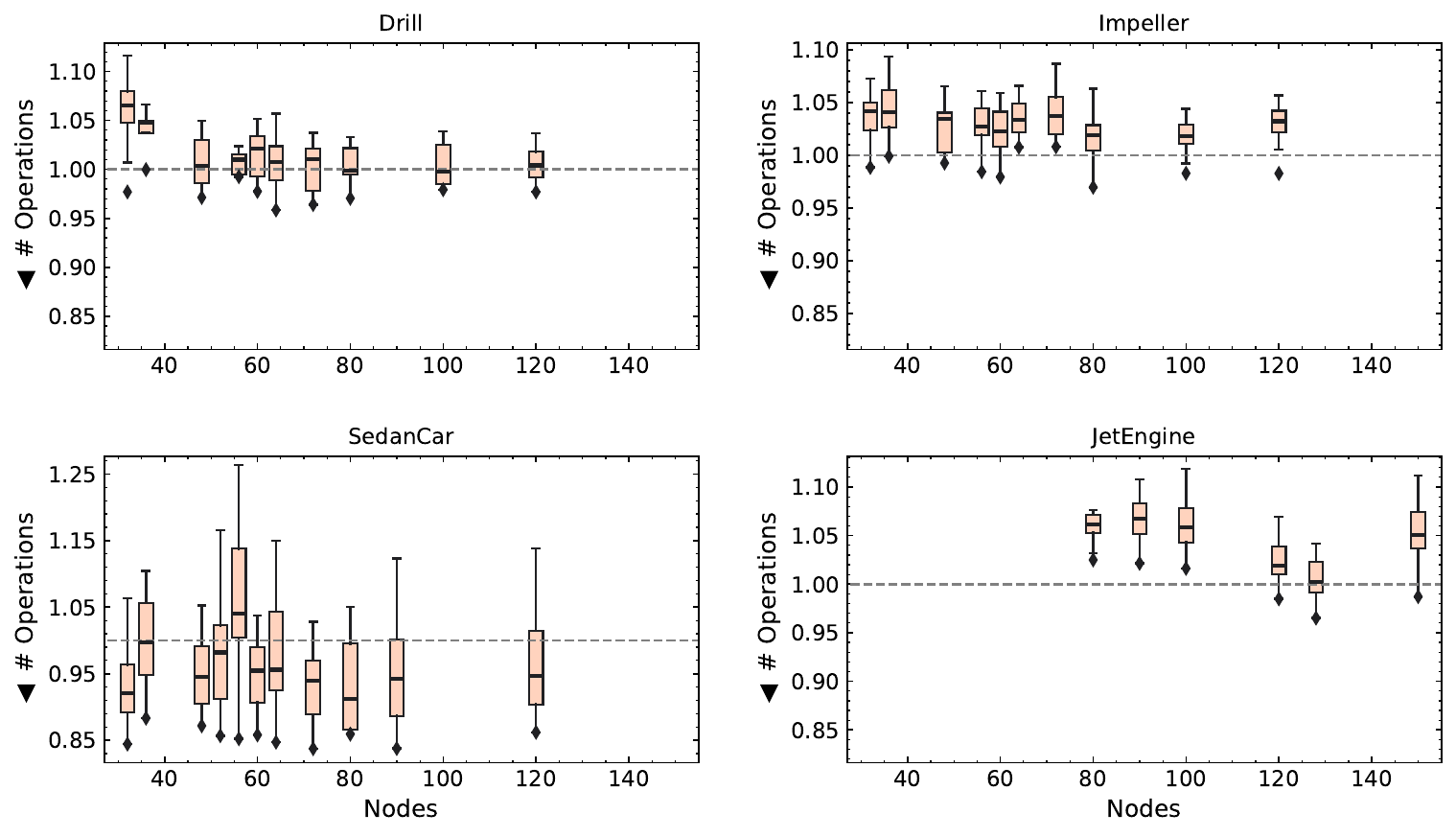}
    \caption{Number of operations computed from symbolic factorization as a function of coarse graph size using quantum-derived partitions. The horizontal dashed line denotes the normalized baseline ($1.0$).}
    \label{fig:obs_boxplots}
\end{figure*}

\begin{figure*}[!htb]
    \centering
    \includegraphics[width=\textwidth]{
    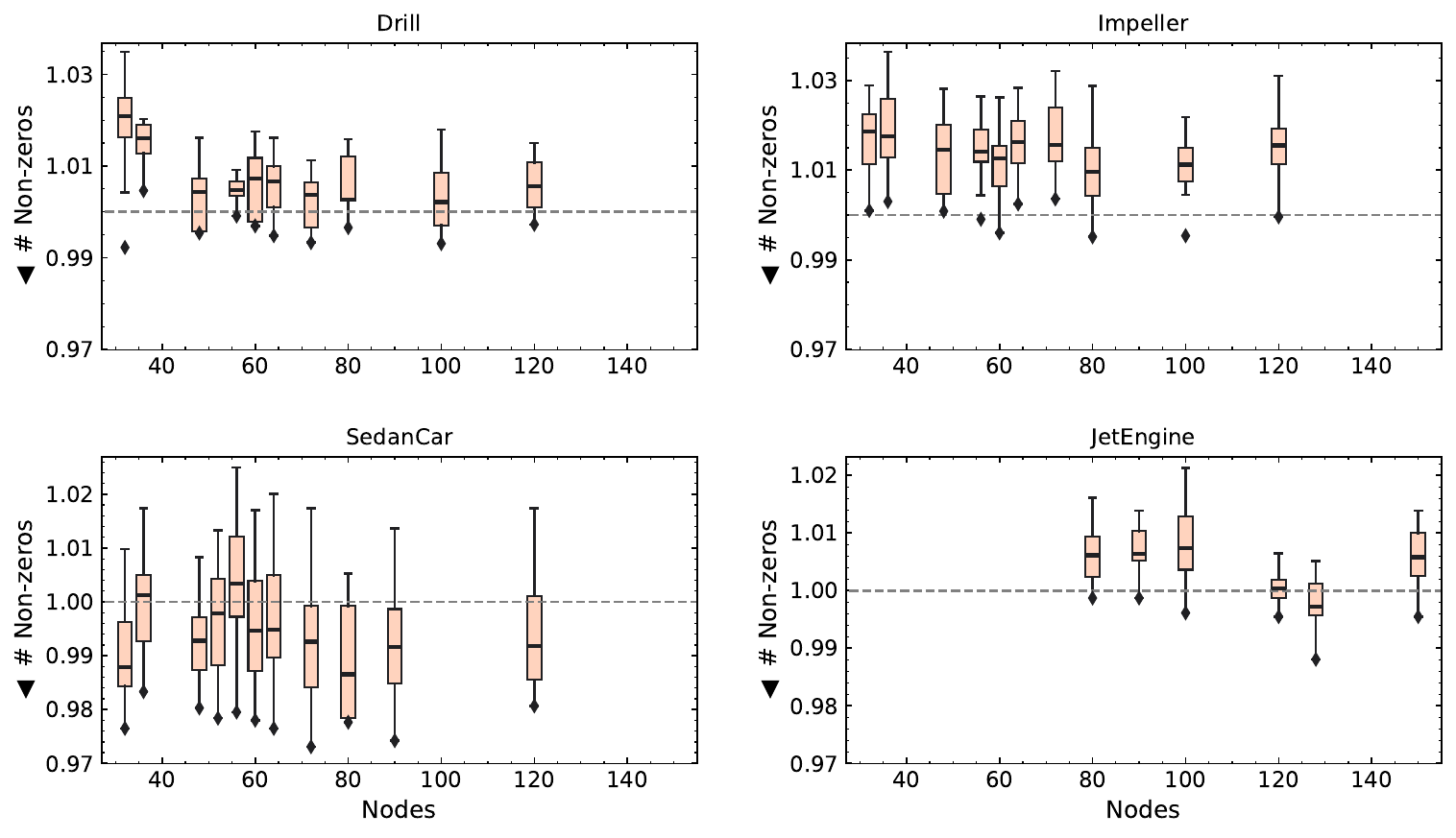}
    \caption{Number of non-zeros computed from symbolic factorization as a function of coarse graph size using quantum-derived partitions. The horizontal dashed line denotes the normalized baseline ($1.0$).}
    \label{fig:nnz_boxplots}
\end{figure*}

\bibliographystyle{IEEEtran}
\bibliography{references}

\end{document}